\documentclass[11pt]{article}
\title{Duality in Quantum Information Geometry}
\author{R. F. Streater, Dept. of Mathematics, King's College London}
\date{9 June 2003}

\newtheorem{theorem}{Theorem}

\newtheorem{definition}[theorem]{Definition}
\begin{document}
\maketitle \setlength{\oddsidemargin}{0in}
\setlength{\evensidemargin}{0in}
\begin{abstract}
Let ${\cal H}$ be a separable Hilbert space. We consider the
manifold ${\cal M}$ consisting of density operators $\rho$ on
${\cal H}$ such that $\rho^p$ is of trace class for some
$p\in(0,1)$. We say $\sigma\in{\cal M}$ is nearby $\rho$ if there
exists $C>1$ such that $C^{-1}\rho\leq \sigma\leq C\rho$. We show
that the space of nearby points to $\rho$ can be furnished with
the two flat connections known as the $(\pm)$-affine structures,
which are dual relative to the {\em BKM} metric. We furnish ${\cal
M}$
with a norm making it into a Banach manifold.
\end{abstract}
\section{Introduction}
\subsection{The problem}
The set of states of a classical or quantum system can be
furnished with the $L^1$-norm relative to a reference measure (or
the trace norm for quantum systems). The associated metric is not
the ideal measure of distance between states. For example, in
infinite dimensions any trace-norm hood of a state $\rho$ of
finite entropy contains a dense set of states of infinite entropy.
These cannot be near $\rho$ in any physical sense. We would also
expect that a viable geometric description of the dynamics of a
macroscopic system would be a path $\{\rho(t),t\geq 0\}$, where
the relative entropy
\begin{equation}
S(\rho(t)|\rho(0)):={\rm Tr}\,\rho(t)(\log\rho(t)-\log\rho(0))
\end{equation}
is finite. This is so because the relative entropy is closely
related to the free energy. These requirements cannot be
controlled by the trace norm. We propose a stronger topology for
the manifold, controlled by a new norm which we call the Araki norm,
as it arises naturally in H. Araki's theory of `expansionals'. This
is sufficient for the relative entropy between a state any any
of its neighbours to be finite, as well as ensuring that the mixture
connection, and the canonical connection, are defined on the manifold.
They are then dual relative to the Bogoliubov-Kubo-Mori metric.
Our condition is far from being necessary for these results, however;
so we can expect that this is not the last word on
dual information geometry.

We start with \S 1.2, a review of our definition
\cite{RFSBog,RFSanal} of the quantum information manifold in
infinite dimensions. This makes use of the theory of relatively
bounded perturbations, and constructs a manifold with a metric and
the $(+1)$-affine structure. In \S~2 we note that the
$(-1)$-affine structure can be defined for a vector subspace of
the tangent space defined there, namely, for nearby states. We say
that a density matrix $\sigma$ is nearby the density matrix $\rho$
if there exists a constant $C>1$ such that
\begin{equation}
C^{-1}\rho\leq\sigma\leq C\rho \label{nearby}
\end{equation}
in the sense of operator inequality.

 In the context of faithful states on
von Neumann algebras, Araki had proved \cite{Arakibounded} that
(\ref{nearby}) is enough to ensure that the relative Hamiltonian of
$\rho$ and $\sigma$ is a bounded element of the algebra, with a
holomorphic property. Our theory is in a more concrete
setting, in which these results are true but not so deep.
By establishing the converse of Araki's
theorem in our simple case, we are able to show that the
$(+1)$-affine structure is retained on this subspace, which therefore
has two flat affine structures.

\subsection{The quantum information manifold} Let ${\cal H}$ be a
separable Hilbert space, and $B({\cal H})$ the W*-algebra of
bounded operators on ${\cal H}$, with norm $\|\,\cdot\,\|$. Recall
\cite{RFSBog} that the quantum information manifold for ${\cal H}$
is constructed as follows. Let ${\cal C}_p$, $0<p<\infty$ be the
vector space of $A\in B({\cal H})$ such that
\begin{equation}
\|A\|_p:={\rm Tr}\left(|A|^p\right)^{1/p}<\infty.
\end{equation}
Then ${\cal C}_p$ is the Schatten class, a Banach space, for
$1\leq p\leq\infty$ and is a complete quasinormed space for
$0<p<1$ \cite{Pietsch}. Let $\Sigma$ be the convex set of (von
Neumann) density operators on ${\cal H}$. Then we defined the
underlying set of the information manifold to be
\begin{equation}
{\cal M}=\bigcup_{0<p<1}{\cal C}_p\cap\Sigma.
\end{equation}
Thus ${\cal M}$ is the set of density operators $\rho$ such that
there exists $p\in(0,1)$ such that $\rho^p$ is of trace-class.
Given $\rho_0\in{\cal M}$ we suggested \cite{RFSBog,RFSanal,RFSepsi}
various norm topologies on ${\cal M}$ for points near $\rho_0$
as follows: write $\rho_0=\exp\{-H_0-\Psi_0\}$, where $H_0\geq 0$,
and let $X$ be
a quadratic form defined on $Q(H_0):={\rm Dom}(H_0^{1/2})$, such that
for some $\epsilon\in[0,1/2]$ we have
\begin{equation}
\|X\|_\epsilon:=\|(H_0+I)^{-1+\epsilon}X(H_0+I)^{-\epsilon}\|<\infty.
\end{equation}
We showed that if this norm is small enough, $H_0+X$ defines a
self-adjoint operator, that
\begin{equation}
\rho_{_X}:=\exp\{-H_0-X-\Psi_X\}\in{\cal M} \label{pert}
\end{equation}
holds, that the generalised expectation $\rho_0\cdot X$ can be
defined, and adjusted to zero, which determines the normalising
constant $\Psi_X$. Then the partition function $Z_X:=e^{\Psi_X}$
was shown to be a Lipschitz continuous function of $X$ in the norm
$\|\,\cdot\,\|_\epsilon$, if $\epsilon=1/2$, and $C^\infty$ in the
Fr\'{e}chet sense, and analytic in the canonical coordinates, if
$0\leq\epsilon<1/2$. We showed that topologised in this way around
each $\rho_0\in{\cal M}$, ${\cal M}$ becomes a Banach manifold.
Each 'hood of a point is a convex set, under the convex structure
$X_1,X_2\mapsto \lambda X_1+(1-\lambda)X_2$, $\lambda\in(0,1)$.
This defines the (+1)-affine structure on ${\cal M}$.

So far, we have been unable to prove that (for some choice of
$\epsilon$) if $\rho,\sigma\in{\cal M}$ then
$\lambda\rho+(1-\lambda)\sigma \in{\cal M}$. Thus, the
$(-1)$-affine structure is missing. In the next section we find a
subspace on which both the $(+1)$-affine structure and the
$(-1)$-affine structures are defined; this turns out to be the
same space as used by Araki in the much more general case of {\em
KMS}-states. It is not complete in any of the epsilon norms;
however, it is complete in a norm arising in Araki's theory.

\section{Nearby states and the $(-1)-$affine structure}
It is convenient to extend the states by scaling, to get finite
weights on $B({\cal H})$. We denote the extended space of states
by $\tilde\Sigma$ and the extended manifold by $\tilde{\cal M}$.
We write a finite weight as $\rho_{_X}=\exp(-H_0-X)$.

We shall also use a more general concept than being nearby:
\begin{definition}
Suppose that there exists $C>1$ and $p\in[0,1)$ such that in
the sense of operator inequalities,
\begin{equation}
C^{-1}\rho^{1+p}\leq\sigma\leq C\rho^{1-p}. \label{pnearby}
\end{equation}
Then we say that $\sigma$ is $p-$nearby $\rho$.
\end{definition}
It is clear that in the case $p=0$ this is an equivalence
relation. This case reduces to (\ref{nearby}), so that $\sigma$
and $\rho$ are {\em nearby} each other.

Let $\rho_0\in\tilde{\Sigma}$ and
suppose that $\rho_{_X}$ is nearby $\rho_0$.
Then Araki \cite{Arakibounded} proved that $X$ is bounded. In our
special setting, this is a corollary to an easy result.
\begin{theorem}
If $\rho_0,\rho_{_X}\in\tilde\Sigma$ and $\rho_{_X}$ is $p-$nearby $\rho$,
then $X$ is an $H_0$-bounded form, with form bound $\leq p$.
\end{theorem}
{\bf PROOF}. The map $X\mapsto \log X$ is an operator monotone
map, which also applies to forms. Taking logs of (\ref{pnearby})
we get as an inequality of forms
\[
-\log C-(1+p)H_0\leq-H_0-X\leq \log C-(1-p)H_0.\]
 This shows that
$Q(X)\subseteq Q(H_0)$, so the forms $X$ and $H_0$ are comparable,
and we may cancel $H_0$ to get
\[
-\log C-pH_0\leq -X\leq \log C+pH_0.\] Thus, X is an $H_0$-bounded
form, with form bound $\leq p$; if $p=0$ we see that $X$ is
bounded by $\log C$.\hfill QED

Suppose now that $\sigma_1$ and $\sigma_2$ are finite weights, and
both are $p-$nearby the finite weight $\rho$. Then
$\lambda\sigma_1+(1-\lambda)\sigma_2$ is $p-$nearby $\rho$. This
follows easily form the definition (\ref{pnearby}). Since any
state is $p-$nearby itself, we see that the set of states that are
$p-$nearby $\rho_0$ form a mixture family. The problem is to find
a norm topology (relative to the $(+1)$-linear structure) such
that all $(-1)$-mixtures lie in the 'hood of the state $\rho$ in
that norm topology.

\section{Araki's analyticity and its converse}
Araki \cite{Arakibounded} showed in the context of {\em KMS}
theory, that if $\rho_{_X}$ is nearby the state $\rho$, then the
map
\begin{eqnarray}
t&\mapsto& \rho^{t}X\rho^{-t}\nonumber\\
&&\mbox{is bounded and holomorphic in the circle } \{t\in{\bf C}:
|t|<1/2\}.\label{araki}
\end{eqnarray}
This suggests the definition of the Araki norm: \begin{definition}
For any $\rho\in{\cal M}$ and any bounded operator $X$, \[
\|X\|_{_A}:=\sup_{0\leq|t|<1/2}\left\{\|\rho^tX\rho^{-t}\|\right\}.\]
\end{definition}
In our context, we prove a converse, to get
\begin{theorem}\label{Araki2}
Suppose that $X\in B({\cal H})$, and that $\rho$ and $\rho_{_X}$
are finite weights. Then (\ref{araki}) holds if and only if
$\rho_{_X}$ is nearby $\rho$.
\end{theorem}
{\bf PROOF}. From Araki's result, we only need to prove the `only
if' direction. We start with the Matsubara expansion
\begin{equation}
\rho_{_X}=\rho+\sum_{n=1}^\infty
\frac{1}{n!}\int_0^1\prod_{j=1}^{n+1}
d\alpha_j\delta\left(\sum_{i=1}^{n+1}\alpha_i-1\right)\rho^
{\alpha_1}X\rho^{\alpha_2}X\ldots X\rho^{\alpha_{n+1}}
\end{equation}
and obtain a formal expansion for the quadratic form
\begin{eqnarray}
\rho^{-1/2}\rho_{_X}\rho^{-1/2}&=&1+\sum_{n=1}^\infty\frac{1}{n!}
\int_0^1\prod
_{j=1}^{n+1}d\alpha_j\delta\left(\sum_{i=1}^{n+1}\alpha_i-1\right)
\left(\rho^{\alpha_1-1/2}X\rho^{1/2-\alpha_1}\right)\nonumber\\
& &\left(\rho^{\alpha_1+\alpha_2-1/2}X\rho^{1/2-\alpha_1-\alpha_2}
\right)\ldots\left(\rho^{1/2-\alpha_{n+1}}X\rho^{\alpha_{n+1}-1/2}\right).
\end{eqnarray}
Each of the components $\pm(\alpha_1+\ldots+\alpha_j-1/2),
\;\;1\leq j \leq n$ lies between $-1/2$ and $1/2$, so by
(\ref{araki}) each term is bounded by $M^n/n!$, where
$M=\|X\|_{_A}$, defined in (\ref{araki}). So the series is norm
convergent to a bounded operator of norm $\leq C:=e^M$. The series
is then known to converge to $\rho^{-1/2}\rho_{_X}\rho^{-1/2}\leq
C$, giving $\rho_X\leq C\rho$. \\In the other direction we
consider the unbounded positive operators $\rho_{_X}^{-1}$ and
$\rho^{-1}$ related by the formal series
\begin{equation}
\rho_{_X}^{-1}=\rho^{-1}+\sum_{n=1}^\infty\frac{1}{n!}\int_0^1\prod
_{j=1}^{n+1}d\alpha_j\delta\left(\sum_{i=1}^{n+1}\alpha_i-1\right)
\rho^{-\alpha_1}X\rho^{-\alpha_2}\ldots X\rho^{-\alpha_{n+1}}.
\end{equation}
Multiplying on the left and right by $\rho^{1/2}$ it becomes a series
of bounded operators:
\begin{eqnarray}
\rho^{1/2}\rho_{_X}^{-1}\rho^{1/2}&=&1+\sum_{n=1}^\infty\frac{1}{n!}\int_0^1
\prod{j=1}^{n+1}d\alpha_j\delta\left(\sum_{i=1}^{n+1}\alpha_i-1\right)
\left(\rho^{1/2-\alpha_1}X\rho^{\alpha_1-1/2}\right)\nonumber\\
& &\left(\rho^{1/2-\alpha_1-\alpha_2}X\rho^{\alpha_1+\alpha_2-1/2}
\right)\ldots\left(\rho^{\alpha_{n+1}-1/2}X\rho^{1/2-\alpha_{n+1}}\right).
\end{eqnarray}
Again, each term is bounded by $M^n/n!$ so we get, as forms
\[
\rho^{1/2}\rho_{_X}^{-1}\rho^{1/2}\leq e^MI, \mbox{ so
}\rho_{_X}^{-1}\leq C\rho^{-1}.\] Now, $A\mapsto A^{-1}$ is an
operator-monotone decreasing map, so we get
\[
\rho_{_X}\geq e^{-M}\rho=C^{-1}\rho. \hspace{1in}\mbox{\bf QED}.\]
From this we see if the the hood of a state $\rho$ be taken to be
the subset of ${\cal M}$ consisting of states $\rho_{_X}$ that are
nearby $\rho_{_0}$, then the tangent space at $\rho$ carries both
the canonical and the mixture affine structures. This follows from
this theorem, and the obvious fact that $(+1)-$mixtures of states
having the property (\ref{araki}) also have that property.

One can immediately see that the norm given by the {\em BKM-}metric is
weaker than the Araki norm:
\[\|X\|_{_{M}}^2=\int_0^1{\rm Tr}\left\{\rho^tX\rho^{1-t}X\right\}dt=
\int_0^1{\rm Tr}\left\{ \rho^{t-1/2}X\rho^{1/2-t}\right\}dt\leq
\|X\|_{_A}^2.\] It can be shown that it is properly weaker; in our
proposed topology, the manifold is not complete in the {\em BKM}
metric, and so also not complete in any epsilon norm, or the
Orlicz norm suggested in \cite{Ray}.
\section{A Banach manifold with canonical-mixture duality}
Let us start with the manifold ${\cal M}$, of \cite{RFSBog} as a
set, but define its topology using the Araki norm rather than one
of the epsilon norms . We extend the manifold by allowing finite
faithful weights as well as faithful states. If $\rho$ is a finite
faithful weight, it can be written as $\rho=\exp\{-H_0\}$. We
allow as points of the hood of $\rho$ all the nearby weights.
These are all finite and faithful. The points in a hood form a
$(-1)-$mixture family. By Theorem(\ref{Araki2}), the family is
parametrized by those bounded operators $X$ such that $\rho^t X
\rho^{-t}$ is bounded and holomorphic in $|t|<\frac{1}{2}$. This
family is convex under the canonical, that is, $(+1)-$affine
structure, since if $X$ and $Y$ obey this condition, so does
$\lambda X+(1-\lambda)Y$. More, the space is complete in the Araki
norm (using the canonical affine structure to define the
uniformity). For, a sequence of functions holomorphic and bounded
in an open disc, which is a Cauchy sequence in the sup norm,
converges to a bounded holomorphic function in the same disc. It
has two flat affine structures, the canonical $(+1)$ and the
mixture, $(-1)$. We now show that they are dual relative to the
{\em BKM} metric.
\begin{theorem}[Duality]
The $(\pm)-$affine structures defined here are dual relative to the
{\em BKM} metric.
\end{theorem}
{\bf PROOF}. By construction, the relative entropy
$S(\rho_{_0}|\rho_{_X})$ between two nearby states $\rho_{_0}$ and
$\rho_{_X}$ is finite, and by the result of \cite{RFSanal}, it is
smooth. By definition, the {\em BKM} metric is half of the second
(Fr\'{e}chet) derivative of this. This is also the metric obtained
as half of the Hessian of the Legendre dual, the other relative
entropy $S(\rho_{_X}|\rho_{_0})$. Then $g$ is the Hessian of
\[
\left(S(\rho_{_0}|\rho_{_X})+S(\rho_{_X}|\rho_{_0}) \right)=-{\rm
Tr}\,\{X(\rho_{_0}-\rho_{_X})\},\] by a simple calculation. Now
$X$ is the displacement in the state along a $(+1)-$geodesic, and
$\rho_{_0}-\rho{_X}$ is minus the displacement along a
$(-1)-$geodesic. If $\rho_{_0} - \rho_{_X}:=\delta\rho$ is small
in the Araki norm, the left-hand side is close to
$g(\delta\rho,\delta\rho)$ and the right-hand side is ${\rm
Tr}\left\{\delta\rho_+\delta\rho_-\right\}$. This is the mixed
representation of the {\em BKM}-metric, and the equation expresses
the duality. \hspace{\fill}{\bf QED}

We may furnish the tangent space of ${\cal M}$ with the Araki
norm, for bounded holomorphic operators $X(t)=\rho^tX\rho^{-t}$ in
the disc $|t|<\frac{1}{2}$. This defines a hood of $\rho$. The
norms in overlapping regions, around $\rho_0$ and $\rho_1$, are
compatible; this requires the equivalence of the norms
\[
sup_{|t|<1/2}\{\|\rho_0^t X\rho_0^{-t}\|\}\hspace{.5in}\mbox{ and }
\hspace{.5in}\sup_{|t|<1/2}\{\|\rho_1^tX\rho_1^{-t}\|\}.\]
This is true, since they are related by conjugation with
the Connes cocycle $\rho_0^t\rho_1^{-t}$, and this is
 bounded and analytic in the disc \cite{Ohyapetz}. In this way, we have a
(smallish) Banach manifold with the dual structure.

We can see that the trace norm on states is weaker than the norm
$\|X\|$ on the space of perturbations, and therefore also weaker
than the Araki norm. Indeed,
\begin{theorem}
If $\rho=\rho_0$ and $\sigma=\rho_{_X}$, then
\[
\|\rho-\sigma\|_1\leq \|X\|.\]
\end{theorem}
{\bf PROOF}. Our equality
\[
S(\rho|\sigma)+S(\sigma|\rho)={\rm Tr}\,\{(\rho-\sigma)X\}\]
shows that (by the quantum Kullback inequality \cite{Hiai},
\[\|\rho-\sigma\|_1^2\leq {\rm Tr}\,\{(\rho-\sigma)X\}\leq
\|\rho-\sigma\|_1\|X\|\] So dividing by $\|\rho-\sigma\|_1$ gives
the theorem. \hspace{\fill}{\bf QED}\\
The Araki norm
\[
\|X\|\leq \|X\|_{_A}:=\sup_{|t|<1/2}\{\|\rho^tX\rho^{-t}\|\}\] is
thus also stronger than the trace norm. The expansional for the
relative entropy shows that all states $\sigma$ in an Araki 'hood
of a point $\rho\in{\cal M}$ has finite relative entropy
$S(\rho|\sigma)$, which was the desired property.

{\bf Acknowledgements}: This work was done during a research visit
to the Science University of Tokyo, Noda. I thank M. Ohya for the
kind invitation, and N. Watanabe and H. Hasegawa for useful
discussions.

\end{document}